\documentclass[%
superscriptaddress,
preprint,
showpacs,
nofootinbib,
 amsmath,amssymb, 
aps, 
eqsecnum
]{revtex4-1}

\usepackage{graphicx}
\usepackage{dcolumn}
\usepackage{bm}

\begin{document}

~
\vspace{1cm}

\title{A no-go theorem for the $\boldsymbol{n}$-twistor description of a massive particle} 
\author{Satoshi Okano}
\email[Author to whom correspondence should be addressed. E-mail:]{ 
okano@phys.cst.nihon-u.ac.jp}
\affiliation{Institute of Quantum Science, College of Science and Technology, 
Nihon University, Chiyoda-ku, Tokyo 101-8308, Japan} 
\affiliation{Department of Quantum Science and Technology, Graduate School of Science and Technology, 
Nihon University, Chiyoda-ku, Tokyo 101-8308, Japan}
\author{Shinichi Deguchi}
\email[E-mail: ]{deguchi@phys.cst.nihon-u.ac.jp}
\affiliation{Institute of Quantum Science, College of Science and Technology, 
Nihon University, Chiyoda-ku, Tokyo 101-8308, Japan} 
\affiliation{Department of Quantum Science and Technology, Graduate School of Science and Technology, 
Nihon University, Chiyoda-ku, Tokyo 101-8308, Japan}
%

\date{\today}

\begin{abstract}
It is proved that the $n$-twistor expression of a particle's four-momentum vector 
reduces, by a unitary transformation, 
to the two-twistor expression for a massive particle or the one-twistor expression for a massless particle. 
Therefore the {\em genuine} $n$-twistor description of a massive particle in 
four-dimensional Minkowski space fails for the case $n\ge3$. 
\end{abstract}

\maketitle

\section{Introduction}

About 40 yeas ago, attempts to describe massive particles and their associated internal symmetries 
were made by Penrose, Perj\'{e}s, and Hughston within the framework of twistor theory 
\cite{Penrose, Perjes1, Perjes2, Perjes3, Perjes4, Hughston}. 
To describe a massive particle in four-dimensional Minkowski space, $\mathbf{M}$, 
they introduced two or more $[^{\;\!}$i.e., $n^{\:\!}(\in \Bbb{N}+1)$]  
independent twistors and their dual twistors
\begin{align}
Z_{i}^{A}=(\omega_{i}^{\alpha}, \pi_{i \dot{\alpha}}) \,, 
\qquad  \bar{Z}^{i}_{A}=(\bar{\pi}^{i}_{\alpha}, \bar{\omega}{}^{i\dot{\alpha}}) \, 
\label{1.1}
\end{align} 
$(A=0,1,2,3^{\:\!} ;^{\:\!} \alpha=0,1^{\:\!} ;^{\:\!} \dot{\alpha}=\dot{0}, \dot{1} )$
distinguished by the index $i$ $(i=1, 2, \ldots n)$.    
Here, $\bar{\pi}^{i}_{\alpha}$ and $\bar{\omega}{}^{i\dot{\alpha}}$ denote the complex conjugates 
of the two-component spinors $\pi_{i \dot{\alpha}}$ and $\omega_{i}^{\alpha}$, respectively: 
$\bar{\pi}^{i}_{\alpha}:=\overline{\pi_{i \dot{\alpha}}}\:\!$,  
$\;\!\bar{\omega}{}^{i\dot{\alpha}}:=\overline{\omega_{i}^{\alpha}}\:\!$.  
The spinors $\omega_{i}^{\alpha}$ and $\pi_{i \dot{\alpha}}$ are related by  
\begin{align}
\omega_{i}^{\alpha}=i z^{\alpha \dot{\alpha}} \pi_{i \dot{\alpha}} \,, 
\label{1.2}
\end{align}
where $z^{\alpha \dot{\alpha}}$ are coordinates of a point in complexified Minkowski space, $\Bbb{C} \mathbf{M}$. 
It was shown in Refs. \cite{Penrose, Perjes1, Perjes2, Perjes3, Perjes4, Hughston} 
that the massive particle system described by $n$ twistors possesses the internal symmetry 
specified by an inhomogeneous extension of $SU(n)$, denoted by $ISU(n)$. 
Penrose, Perj\'{e}s, and Hughston proposed the idea of identifying the $SU(2)$ $[^{\:\!}$or $ISU(2)$]  
symmetry in the two-twistor system with the symmetry for leptons, 
and the $SU(3)$ $[^{\:\!}$or $ISU(3)$] symmetry in the three-twistor system with the symmetry for hadrons.

Long after Penrose, Perj\'{e}s, and Hughston made their attempts, 
Lagrangian mechanics of a massive spinning particle in $\mathbf{M}$ has been formulated in terms of twistors, 
and it has been studied until quite recently 
\cite{FedZim, BALE, AFLM, FFLM, AIL, MRT, FedLuk, AFIL, MRT2, RT, DegOka}. 
In almost all these studies 
\cite{FedZim, BALE, AFLM, FFLM, AIL, MRT, FedLuk, AFIL, MRT2, DegOka}, 
only the two-twistor description of a massive particle is conventionally adopted   
without clarifying the reason why the $n(\ge 3)$-twistor description is not employed. 
Under such circumstances, Routh and Townsend showed that only the two-twistor formulation 
can successfully describe a massive particle in $\mathbf{M}$ \cite{RT}. 
(See also the note added in Sec. 3.)

As can be seen in Refs. \cite{Penrose, Perjes1, Perjes2, Perjes3, Perjes4, Hughston},  
the $n$-twistor expression of a particle's four-momentum vector is given by 
\begin{align}
p_{\alpha\dot{\alpha}}=\sum_{i=1}^{n} \bar{\pi}^{i}_{\alpha} \pi_{i \dot{\alpha}}  
\equiv \bar{\pi}^{i}_{\alpha} \pi_{i \dot{\alpha}} \,. 
\label{1.3}
\end{align}
(The expression (\ref{1.3}) has recently been exploited 
to realize massive representations of the Poincar\'{e} algebra \cite{CJM}. 
In this study, the massive representations are actually considered only in the case $n=2$.) 
By using Eq. (\ref{1.3}), the squared mass $m^{2} =p_{\alpha\dot{\alpha}} p^{\alpha\dot{\alpha}}$ can be written as 
\begin{align}
m^{2} =\bar{\pi}^{i}_{\alpha} \pi_{i \dot{\alpha}} \bar{\pi}^{j \alpha} \pi_{j}^{\dot{\alpha}}, 
\label{1.4}
\end{align}
where $\bar{\pi}^{i \alpha}:=\epsilon^{\alpha\beta} \bar{\pi}^{i}_{\beta}$ 
and $\pi_{i}^{\dot{\alpha}}:=\epsilon^{\dot{\alpha} \dot{\beta}} \pi_{i \dot{\beta}}$ 
($\epsilon^{01}=\epsilon^{\dot{0} \dot{1}}=1$).    
In Lagrangian mechanics mentioned above, Eq. (\ref{1.4}) with $n=2$, 
or its equivalent expression  
\begin{subequations}
\label{1.5}
\begin{align}
\epsilon^{ij} \pi_{i \dot{\alpha}} \pi_{j}^{\dot{\alpha}} -\sqrt{2} \:\! m e^{i\varphi} &=0 \,, 
\label{1.5a}
\\
\epsilon_{ij} \bar{\pi}^{i}_{\alpha} \bar{\pi}^{j \alpha} -\sqrt{2} \:\! m e^{-i\varphi} &=0 
\label{1.5b}
\end{align}
\end{subequations}
($\epsilon^{12}=\epsilon_{12}=1$) with a real parameter $\varphi$ \cite{FedLuk, AFIL, DegOka},  
is incorporated into a generalization of the Shirafuji action \cite{Shirafuji} with the aid of Lagrange multipliers.

Now, we present the following theorem: 

\vspace{7mm}

\noindent
{\bf Theorem:}~ 
The $n$-twistor expression given in Eq. (\ref{1.3}) reduces to the two-twistor expression 
\begin{align}
p_{\alpha\dot{\alpha}} &=\Bar{\Tilde{\pi}}{}^{1}_{\alpha} \tilde{\pi}_{1 \dot{\alpha}} 
+\Bar{\Tilde{\pi}}{}^{2}_{\alpha} \tilde{\pi}_{2 \dot{\alpha}} \,, 
\qquad  \tilde{\pi}_{1 \dot{\alpha}}  \tilde{\pi}_{2}^{\dot{\alpha}} \neq 0  
\label{1.6} 
\end{align} 
or the one-twistor expression 
\begin{align}
p_{\alpha\dot{\alpha}} &=r \Bar{\Tilde{\pi}}{}^{1}_{\alpha} \tilde{\pi}_{1 \dot{\alpha}} \,, 
\label{1.7}
\end{align}
where $\tilde{\pi}_{i \dot{\alpha}}:=U_{i}{}^{j} \pi_{j \dot{\alpha}}{}^{\:\!}$,  
$\Bar{\Tilde{\pi}}{}^{i}_{\alpha} \big(:=\overline{\tilde{\pi}_{i \dot{\alpha}}}\, \big)
=\bar{\pi}^{j}_{\alpha} U^{\dagger}{}_{j}{}^{i}$, and 
$r$ is a real constant. Here, $U_{i}{}^{j}$ are the entries of an $n \times n$ unitary matrix $U$. 

\vspace{7mm}

Hence the $n$-twistor system eventually turns out to be a two-twistor system representing a massive particle 
or a one-twistor system representing a massless particle. 
The purpose of the present paper is to prove this theorem. 
The theorem leads to the fact that the {\em genuine} $n(\ge3)$-twistor description 
of a massive particle in $\mathbf{M}$ fails owing to the reduction from Eq. (\ref{1.3}) to Eq. (\ref{1.6}) or Eq. (\ref{1.7}) 
caused by a unitary transformation. 
For this reason, the above-mentioned idea for the $SU(3)$ $[^{\:\!}$or $ISU(3)$] symmetry cannot be accepted. 
In this sense, the theorem given here can be said to be a no-go theorem.
Also, the theorem justifies the fact that in the context of a four-dimensional Minkowski background, 
only the two-twistor description (i.e., the case $n=2$) 
has been considered in Lagrangian mechanics of a massive spinning particle 
formulated in terms of twistors.

This paper is organized as follows. 
In Sec. II, we prove the theorem using a lemma provided there.  
Section III is devoted to remarks.

\section{A proof of the theorem}

We first provide a lemma necessary to prove the theorem. 

\vspace{7mm}

\noindent
{\bf Lemma:}~ 
Let $A$ be an $n \times n$ complex antisymmetric matrix, satisfying $A^{\rm T} =-A$.  
Then $A$ can be transformed into its normal form, $\tilde{A}$, according to 
\begin{align}
\tilde{A}=UAU^{\rm T} , 
\label{2.1}
\end{align}
where $U$ is an $n \times n$ unitary matrix, satisfying $U^{\dagger}=U^{-1}$.  
If $n$ is even, then the normal form $\tilde{A}$ is given by 
\begin{align}
\tilde{A}=
\left( 
\begin{array}{ccccccc}
0 & \sqrt{a_1} & 0 & 0  & \cdots & 0 & 0  \\
-\sqrt{a_1} & 0 & 0 & 0 & \cdots & 0 & 0   \\ 
0 & 0 & 0 & \sqrt{a_2} & \cdots & 0 & 0  \\
0 & 0 & -\sqrt{a_2} & 0 & \cdots & 0 & 0  \\
\vdots & \vdots & \vdots & \vdots & \ddots & \vdots & \vdots  \\
0 & 0 & 0 & 0 & \cdots & 0 & \sqrt{a_{n/2}} \\
0 & 0 & 0 & 0 & \cdots & -\sqrt{a_{n/2}} & 0 \\
\end{array} 
\right) ,
\label{2.2}
\end{align} 
and if $n$ is odd, then $\tilde{A}$ is given by 
\begin{align}
\tilde{A}=
\left( 
\begin{array}{cccccccc}
0 & \sqrt{a_1} & 0 & 0  & \cdots & 0 & 0 \;\, & 0 \:\, \\
-\sqrt{a_1} & 0 & 0 & 0 & \cdots & 0 & 0 \;\, & 0 \:\, \\ 
0 & 0 & 0 & \sqrt{a_2} & \cdots & 0 & 0 \;\, & 0 \:\, \\
0 & 0 & -\sqrt{a_2} & 0 & \cdots & 0 & 0 \;\, & 0 \:\, \\
\vdots & \vdots & \vdots & \vdots & \ddots & \vdots & \vdots \;\, & \vdots \:\, \\
0 & 0 & 0 & 0 & \cdots & 0 & \sqrt{a_{(n-1)/2}} \;\, & 0 \:\, \\
0 & 0 & 0 & 0 & \cdots & -\sqrt{a_{(n-1)/2}} & 0 \;\, & 0 \:\, \\ 
0 & 0 & 0 & 0 & \cdots & 0 & 0 \;\, & 0 \:\, \\ 
\end{array} 
\right) . 
\label{2.3}
\end{align} 
Here, $a_{1}, a_{2}, \ldots, a_{n/2}$ $[^{\:\!}$or $a_{(n-1)/2}$] are eigenvalues of the Hermitian matrix $AA^{\dagger\:\!}$;    
hence it follows that these eigenvalues are non-negative real numbers. 

\vspace{7mm}

In this paper, we do not give the proof of this lemma, 
because the proof can be seen in Refs. \cite{Youla, Zumino, Becker, Napoly}.  

\vspace{7mm}

\noindent
{\bf Proof of the theorem:}~ 
In order to prove the theorem, let us consider the $n\times n$ complex matrix $\varPi$ consisting of the elements 
\begin{align}
\varPi_{ij}:=\pi_{i \dot{\alpha}} \pi_{j}^{\dot{\alpha}} .
\label{2.4}
\end{align}
Because $\pi_{i \dot{\alpha}} \pi_{j}^{\dot{\alpha}} =-\pi_{i}^{\dot{\alpha}} \pi_{j \dot{\alpha}}$ holds, 
$\varPi$ turns out to be antisymmetric. 
According to the lemma, the matrix $\varPi$ can be transformed into its normal form 
\begin{align}
\tilde{\varPi}=U \varPi U^{\rm T} 
\label{2.5}
\end{align}
by means of an appropriate $n \times n$ unitary matrix $U=(U_{i}{}^{j})$. 
Expressions corresponding to Eqs. (\ref{2.2}) and (\ref{2.3}) are concisely given by
\begin{subequations}
\label{2.6}
\begin{align}
~\qquad \quad \quad \tilde{\varPi}_{2r-1,\:\! j} &=\delta_{2r,\:\! j} \tilde{\varPi}_{2r-1, 2r} \,, 
\label{2.6a}
\\
~\qquad \quad \quad \tilde{\varPi}_{2r,\:\! j} &=\delta_{2r,\:\! j+1} \tilde{\varPi}_{2r, 2r-1} \,,
\label{2.6b}
\\
& \!\!\!\!\! \!\!\!\!\! \!\!\!\!\! \!\!\!\!\! \!\!\!\!\! 
\!\!\!\!\! \!\!\!\!\! \!\!\!\!\! \!\!\!\!\! 
\begin{cases}
\; r =1,2, \ldots, n/2 \,, & \quad \text{for $n$ even}\,, \\
\; r =1,2, \ldots, (n+1)/2 \,, & \quad \text{for $n$ odd}\,, 
\end{cases}
\notag
\end{align}
\end{subequations}
where $\tilde{\varPi}_{2r-1, 2r}\big(=-\tilde{\varPi}_{2r, 2r-1}\big)$ is the square root of an eigenvalue of $\varPi \varPi^{\dagger}$,  
so that it is a non-negative real number.

Substituting Eq. (2.4) into Eq. (2.5),   
we can express the elements of $\tilde{\varPi}$ as 
\begin{align}
\tilde{\varPi}_{ij} =\tilde{\pi}_{i \dot{\alpha}} \tilde{\pi}_{j}^{\dot{\alpha}} ,
\label{2.7}
\end{align}
with the two-component spinor 
\begin{align}
\tilde{\pi}_{i \dot{\alpha}}:=U_{i}{}^{j} \pi_{j \dot{\alpha}} \,.
\label{2.8}
\end{align}
Since $U$ is unitary and hence invertible, Eq. (\ref{2.8}) can be inversely solved as 
$\pi_{i \dot{\alpha}}=U^{\dagger}{}_{i}{}^{j} \tilde{\pi}_{j \dot{\alpha}}$. 
Substituting this and its complex conjugate into Eq. (\ref{1.3}), we have 
\begin{align}
p_{\alpha\dot{\alpha}}=\sum_{i=1}^{n} \Bar{\Tilde{\pi}}{}^{i}_{\alpha} \tilde{\pi}_{i \dot{\alpha}}  
\equiv \Bar{\Tilde{\pi}}{}^{i}_{\alpha} \tilde{\pi}_{i \dot{\alpha}}  \,. 
\label{2.9}
\end{align}
From Eqs. (\ref{2.6a}) and (\ref{2.7}), 
we see $\tilde{\pi}_{1 \dot{\alpha}} \tilde{\pi}_{k}^{\dot{\alpha}}=0$ ($k=3,4,\ldots, n$).  
This implies that $\tilde{\pi}_{k \dot{\alpha}}$ ($k=3,4,\ldots, n$) 
is proportional to $\tilde{\pi}_{1 \dot{\alpha}}\:\!$, i.e., 
\begin{align}
\tilde{\pi}_{k \dot{\alpha}} =\rho_{k1} \tilde{\pi}_{1 \dot{\alpha}} \,, 
\qquad \rho_{k1} \in \Bbb{C} \,. 
\label{2.10}
\end{align}
Substituting Eq. (\ref{2.10}) into Eq. (\ref{2.7}) and noting the property  
$\pi_{i \dot{\alpha}} \pi_{i}^{\dot{\alpha}}=0$ (no sum with respect to $i^{}$), we obtain  
\begin{align}
\tilde{\varPi}_{kl} 
=\tilde{\pi}_{k \dot{\alpha}} \tilde{\pi}_{l}^{\dot{\alpha}} 
=\rho_{k1} \rho_{l1} \tilde{\pi}_{1 \dot{\alpha}} \tilde{\pi}_{1}^{\dot{\alpha}} =0 \,, 
\qquad 
k, l=3,4,\ldots,n \,. 
\label{2.11}
\end{align}
By using Eq. (\ref{2.10}), 
$\tilde{\varPi}_{2k}=\tilde{\pi}_{2 \dot{\alpha}} \tilde{\pi}_{k}^{\dot{\alpha}}$ ($k=3,4,\ldots, n$) 
can be written as 
\begin{align}
\tilde{\varPi}_{2k} 
=\rho_{k1} \tilde{\pi}_{2 \dot{\alpha}}  \tilde{\pi}_{1}^{\dot{\alpha}} =\rho_{k1} \tilde{\varPi}_{21} \,.  
\label{2.12} 
\end{align}
Equations (\ref{2.6b}) and (\ref{2.12}) together give 
\begin{align}
\rho_{k1} \tilde{\varPi}_{21}=0 \,, 
\qquad 
k=3,4,\ldots,n \,, 
\label{2.13}
\end{align}
with which we consider the following two cases:  
(a) $\tilde{\varPi}_{21} \neq 0\,$ and (b) $\tilde{\varPi}_{21} =0$.

\subsection{Case (a)}

If $\tilde{\varPi}_{21} \neq 0$, then it follows that $\rho_{k1}=0$ for any arbitrary $k$. 
Accordingly, Eq. (\ref{2.10}) becomes 
\begin{align}
\tilde{\pi}_{k \dot{\alpha}}=0 \,, 
\qquad 
k=3,4,\ldots,n \,. 
\label{2.14}
\end{align}
Substituting Eq. (\ref{2.14}) into Eq. (2.9), we have 
\begin{align}
p_{\alpha\dot{\alpha}} &=\Bar{\Tilde{\pi}}{}^{1}_{\alpha} \tilde{\pi}_{1 \dot{\alpha}} 
+\Bar{\Tilde{\pi}}{}^{2}_{\alpha} \tilde{\pi}_{2 \dot{\alpha}} \,. 
\label{2.15} 
\end{align} 
Here, $\tilde{\pi}_{1 \dot{\alpha}}  \tilde{\pi}_{2}^{\dot{\alpha}} 
\big(=-\tilde{\varPi}_{21} \big) \neq 0$ has already been assumed.

\subsection{Case (b)}

In case (b), each of the $\rho_{k1}$ does not need to vanish. 
Combining $\tilde{\varPi}_{21}=0$, Eq. (\ref{2.11}), and 
the $\tilde{\varPi}_{1k}=\tilde{\varPi}_{2k}=0$ ($k=3,4,\ldots, n$) 
included in Eq. (\ref{2.6}) together, we ultimately have 
$\tilde{\varPi}_{ij} =0$ or, equivalently, $\tilde{\pi}_{i \dot{\alpha}} \tilde{\pi}_{j}^{\dot{\alpha}}=0$. 
This leads to 
\begin{align}
\tilde{\pi}_{i \dot{\alpha}} =\rho_{i1} \tilde{\pi}_{1 \dot{\alpha}} \,, 
\qquad \rho_{i1} \in \Bbb{C} \,, \; \rho_{11}=1 \,. 
\label{2.16}
\end{align}
Substituting Eq. (\ref{2.16}) into Eq. (\ref{2.9}), we obtain 
\begin{align}
p_{\alpha\dot{\alpha}} &=r \Bar{\Tilde{\pi}}{}^{1}_{\alpha} \tilde{\pi}_{1 \dot{\alpha}} \,, 
\qquad 
r:=\sum_{i=1}^{n} |\rho_{i1}|^{2} \in \Bbb{R}\,.
\label{2.17}
\end{align}
We have indeed derived Eqs. (\ref{1.6}) and (\ref{1.7}) by investigating the two cases (a) and (b).  
Thus the proof of the theorem is complete. \hspace{\fill} $\blacksquare$ 

\vspace{7mm} 

In case (a), the squared mass $m^{2} =p_{\alpha\dot{\alpha}} p^{\alpha\dot{\alpha}}$ 
becomes $m^{2} =2\big|\tilde{\varPi}_{21} \big|{}^{2} \neq 0$ and it follows that 
the corresponding particle is massive, while, in case (b), the squared mass becomes 
$m^{2} =r^{2} \big|\tilde{\varPi}_{11} \big|{}^{2} =0$ and it follows that 
the corresponding particle is massless.

Now we consider the new twistor 
$\tilde{Z}{}_{i}^{A}:=U_{i}{}^{j} Z_{j}^{A}
=(\tilde{\omega}_{i}^{\alpha}, \tilde{\pi}_{i \dot{\alpha}})$, 
with $\tilde{\omega}_{i}^{\alpha}:=i z^{\alpha \dot{\alpha}} \tilde{\pi}_{i \dot{\alpha}}$. 
In case (a), Eq. (\ref{2.14}) gives $\tilde{Z}{}_{k}^{A}=0$ ($k=3,4,\ldots, n$), and  
we see that the $n$-twistor system turns out to be the two-twistor system described by  
$\big( \tilde{Z}{}_{1}^{A},  \tilde{Z}{}_{2}^{A} \big)$. 
In case (b), Eq. (\ref{2.16}) gives $\tilde{Z}{}_{i}^{A}=\rho_{i1} \tilde{Z}{}_{1}^{A}$, and 
we see that all the twistors $\tilde{Z}{}_{i}^{A}$ correspond to a single projective twistor 
defined as the proportionality class 
$\big[ \tilde{Z}{}^{A}_{1} \big] :=\big\{ \rho \tilde{Z}{}^{A}_{1} \big|\, \rho \in \Bbb{C}\setminus\{0\} \big\}$. 
Hence, in case (b), the $n$-twistor system is described by $\big[ \tilde{Z}{}^{A}_{1} \big]$  
and turns out to be essentially a one-twistor system. 
(This statement is consistent with the expression in Eq. (\ref{2.17}).) 
As is well known in twistor theory, a projective twistor precisely specifies the configuration of a massless particle.   
From this fact, we see again that in case (b), the $n$-twistor system represents a massless particle.

\section{Remarks}

It is now clear that in case (a), the $SU(n)$ $[^{\:\!}$or $ISU(n)$] symmetry of the $n$-twistor system 
reduces to the $SU(2)$ $[^{\:\!}$or $ISU(2)$] symmetry of a two-twistor system for a massive particle, while,  
in case (b), the $SU(n)$ $[^{\:\!}$or $ISU(n)$] symmetry is realized in a one-twistor system for a massless particle. 
Hence, the $SU(n)$ $[^{\:\!}$or $ISU(n)$] symmetry in the case $n \geq 3$ cannot be identified with the internal symmetry 
of a massive physical system consisting of, e.g., hadorns. 
For this reason, the idea proposed by Penrose, Perj\'{e}s, and Hughston fails in the case $n\ge3$. 
Of course, there still remains a possibility that the $SU(n)$ $[^{\:\!}$or $ISU(n)$] symmetry 
will be identified with the internal symmetry of a massless system.

As can be seen from the theorem proved above, 
the case $n=2$ is the only case in which a massive particle in $\mathbf{M}$ can be described. 
In fact, Lagrangian mechanics of a massive spinning particle in $\mathbf{M}$ has been successfully formulated in 
Refs. \cite{FedZim, BALE, AFLM, FFLM, AIL, MRT, FedLuk, AFIL, MRT2, RT, DegOka} by using 
two twistors. In the respective formulations, 
the $SU(2)$ symmetry in two-twistor system is maintained  
in the action of a massive spinning particle. 
It was shown in Ref. \cite{DegOka} that this $SU(2)$ symmetry is a symmetry realized in 
the particle-antiparticle doublet, not in 
the lepton doublet consisting of, e.g., the electron and the electron-neutrino. 
Therefore the idea proposed by Penrose, Perj\'{e}s, and Hughston is not valid in 
the present Lagrangian mechanics based on twistor theory.  
In addition to the $SU(2)$ symmetry, the action possesses the $U(1)$ symmetry 
due to a common phase transformation of two twistors.  
It was pointed out that this symmetry is identified as 
a symmetry leading to chirality conservation \cite{DegOka}. 

\vspace{3mm}

\noindent
{\bf Note added:}~ 
After completing an earlier version of this paper, the authors were informed that 
the same result concerning values of $n$ was obtained by Routh and Townsend 
using an inequality \cite{RT}. 
Because this inequality is derived by considering the phase space dimension and 
appropriate constraints including the spin-shell constraints, 
the result being obtained may be understood to be depending on a specific model.
In contrast, our proof of the theorem is purely linear algebraic  
and is independent of Lagrangian mechanics.  

The authors were also informed that the twistor description of a massive particle in anti-de Sitter space 
has been performed with the use of more than one twistor \cite{CRZ, Cederwall, ABT}. 
In this context, it would be of considerable interest to clarify the necessary number of twistors 
by taking a similar linear algebraic approach. 
We hope to address this issue in the future.

\section*{Acknowledgments}

The authors would like to thank the anonymous reviewer for his/her useful information 
on the twistor description of a particle in anti-de Sitter space.

\end{document}